\documentclass[twocolumn,english,prl,showpacs]{revtex4}
\usepackage[T1]{fontenc}
\usepackage[latin9]{inputenc}
\usepackage{textcomp}
\usepackage{amsmath}
\usepackage{amssymb}
\usepackage{graphicx}
\usepackage{esint}

\makeatletter
\@ifundefined{textcolor}{}
{%
 \definecolor{BLACK}{gray}{0}
 \definecolor{WHITE}{gray}{1}
 \definecolor{RED}{rgb}{1,0,0}
 \definecolor{GREEN}{rgb}{0,1,0}
 \definecolor{BLUE}{rgb}{0,0,1}
 \definecolor{CYAN}{cmyk}{1,0,0,0}
 \definecolor{MAGENTA}{cmyk}{0,1,0,0}
 \definecolor{YELLOW}{cmyk}{0,0,1,0}
 }

\makeatother

\usepackage{babel}
\begin{document}

\title{Transverse coherent transition radiation for diagnosis of modulated
proton bunches}

\author{A. Pukhov and T.Tueckmantel}

\affiliation{Institut fuer Theoretische Physik I, Universitaet Duesseldorf, 40225
Germany}
\begin{abstract}
Transverse coherent transition radiation (TCTR) emitted by a relativistic
particle bunch traversing a conducting surface is analyzed. The bunch
emits dipole-like radiation in the direction transverse to the bunch
axis when the beam radius is smaller than the radiation wavelength.
The radiation wavelength is defined by the longitudinal structure
of the particle bunch. The particular case of proton bunches modified
by propagation in plasma, but still carrying an unmodulated current
is considered. Radius-modulated bunches with a constant current emit
axially symmetric radiation. Hosed bunches emit antisymmetric radiation
in the plane of hosing. The TCTR field amplitude may reach 100 kV/m
for the existing proton bunches.
\end{abstract}


\maketitle
Coherent transition radiation (CTR) is one of the most common techniques
used for diagnosis of a longitudinal structure of charged particles
bunches \cite{rosenzweig,CTR2,CTR3,CTR4,CTR5}. The method particularly
demonstrated its power to characterize accelerated electron bunches
in laser-plasma experiments \cite{leemans-2003,shroeder 2004,tilborg-1}.
An elementary charge propagating through a medium with a particular
dielectric permittivity is dressed by a field matched to that medium.
When the charge traverses a sharp boundary of two media with different
permittivities, its field must be adjusted. The unmatched field can
be radiated. The strongest radiation is observed when a charge passes
a boundary between a conductor and vacuum. A point-like relativistic
charge with the relativistic factor $\gamma$ emits a radiation burst
that is collimated within a cone with the opening angle $\theta\approx1/\gamma$
around the axis, although the emission is exactly zero in the propagation
direction itself. The radiation is broadband. A bunch of particles
can emit this radiation coherently at the wavelength comparable with
its longitudinal structure. 

Recently, a concept of proton bunch-driven plasma wake field accelerator
has been put forward \cite{smpwa,pwa,smpwa theory}. In this concept,
a long proton bunch is sent through plasma where it undergoes self-modulation
at the plasma wave period and excites a strong resonant wake field.
A test experiment is in preparation at CERN. One of the experimental
challenges will be the detection and characterization of the proton
bunch modulation after it exits the plasma cell.

The nature of the proton bunch modulation is such that the proton
bunch radius is modulated, but the total bunch current remains the
same in each cross-section. For this reason, there will be no signatures
of the proton bunch modulation in the forward coherent transition
radiation. The classic forward CTR is cast useless in this case. Moreover,
it is important in the experiment to distinguish between the axisymmetric
modulation mode when the radius of the proton bunch is changing periodically
\cite{smpwa theory} and the possible hosing mode when the proton
bunch centroid oscillates periodically in the transverse plain \cite{hosing}. 

Below we show that the transverse coherent transition radiation (TCTR)
does contain the signature of the bunch modulation and allows to distinguish
between the axisymmetric modulation mode and the hosing. The TCTR
is emitted perpendicularly to the particle bunch propagation direction
and its amplitude does not depend on the particles $\gamma-$factor
as soon as it is large enough.

\section{Origin of transverse transition radiation}

\begin{figure}
\includegraphics[width=0.9\columnwidth]{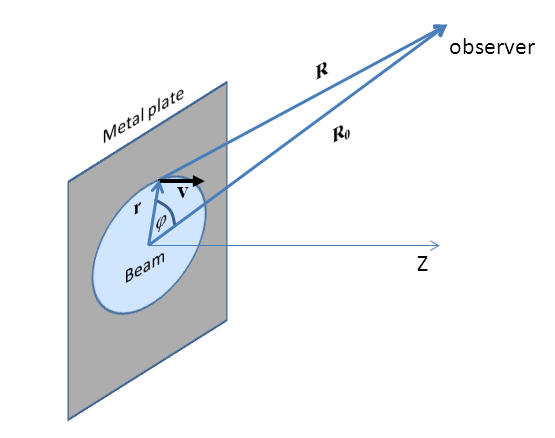}

\caption{\label{fig:geometry} Geometry of the transverse coherent transition
radiation. A charged particle bunch exits a conductor plate normally
to the surface. The observer looks at the emitted radiation from side
to the beam, in the plane of the surface.}
\end{figure}

Let us consider a transition radiation emitted by a particle bunch
as it traverses normally a conductor plate. The interaction geometry
is illustrated in Fig. \ref{fig:geometry}. When an elementary charge
$dq$ exits from the conducting plate in the normal direction with
the velocity $\mathbf{v}$, the radiated field is given by the formula
(63.8) from the Landau textbook \cite{landau}:
\begin{eqnarray*}
\mathbf{dE} & = & \frac{dq}{c^{2}\left(R-\frac{\mathbf{Rv}}{c}\right)^{3}}\mathbf{R}\times\left[\left(\mathbf{R}-\frac{\mathbf{v}}{c}R\right)\times\frac{d\mathbf{v}}{dt'}\right]
\end{eqnarray*}
\begin{eqnarray}
 & + & \frac{dq}{c^{2}\left(R+\frac{\mathbf{Rv}}{c}\right)^{3}}\mathbf{R}\times\left[\left(\mathbf{R}+\frac{\mathbf{v}}{c}R\right)\times\frac{d\mathbf{v}}{dt'}\right]\label{eq:lienard-wiechert}
\end{eqnarray}
where $t'$ is the retarded time so that 
\begin{equation}
t'+R(t')/c=t.\label{eq:retarded}
\end{equation}
The second term in Eq. \eqref{eq:lienard-wiechert} is generated by
the image of the physical charge in the conducting plate.

When the elementary charge is inside the metal plate, its field is
completely screened. Thus, the current is created abruptly when the
charge exits into the free space. We can write for the velocity $\mathbf{v}(t')=\mathbf{v}_{0}\Theta(t'-t_{\mathrm{e}})$,
where $t_{\mathrm{e}}$ is the time the charge exits into vacuum and
$\Theta(t)$ is the Heaviside step function. 

The denominators in Eq. \eqref{eq:transverse} suggest that a point-like
charge emits the strongest field at the angle $\theta\approx1/\gamma$
around the propagation direction. Yet, we will be not interested in
the emission in this direction, because it does not contain any coherent
information about the bunch modulation. Rather, we are looking in
the transverse plane, along the surface of the conducting plate. It
is the dipole-like radiation of the particle bunch .

For the radiation emitted in the transverse plane, Eq. \eqref{eq:transverse}
becomes especially simple, because $\mathbf{v\perp R}$:

\begin{equation}
\mathrm{\mathit{\mathbf{dE}}}=-\frac{2\mathbf{v_{0}}dq}{c^{2}R}\delta(t'-t_{\mathrm{e}})\label{eq:transverse}
\end{equation}
This is essentially the dipole radiation formula. For a continuous
bunch, $dq=en(t,\mathbf{r})v_{0}dSdt,$ where $dS=rdrd\varphi$ is
the area element in the bunch cross-section and $n$ is the local
particle density. The field is polarized in the bunch propagation
direction and normal to the metal plate surface. The observed radiated
field is thus
\begin{equation}
E(t,R_{0})=-\frac{2ev_{0}^{2}}{c^{2}R_{0}}\iint n(t',\mathbf{r})dS\label{eq:transverse-1}
\end{equation}
Where we have assumed the observer being far away from the source
and set $R\approx R_{0}$ in the denominator. At the same time, the
retarded time is

\begin{equation}
t'=t-\frac{R(t')}{c}\approx t-\frac{R_{0}}{c}+\frac{r}{c}\cos\varphi.\label{eq:retarded-1}
\end{equation}

Now we are ready to apply the formula \eqref{eq:transverse-1} to
specific particle bunches. We are interested primarily in proton bunches
that have undergone self-modulation or hosing after they propagated
in plasma. The peculiarity of these proton bunches is that the current
is not modulated. Only the bunch radius may be modulated or the centroid
position is changing (hosing). We show that self-modulation and hosing
result in TCTR with different signatures that can be experimentally
observed.

\section{TCTR emitted by modulated axially symmetric bunches with constant
current}

In this section we consider TCTR of a axially symmetric particle bunch
whose radius is modulated. For simplicity, we assume the Gaussian
transverse bunch density distribution
\begin{equation}
n(t',\mathbf{r},z)=n_{0}\frac{\sigma_{0}^{2}}{\sigma^{2}(t',z)}e^{-r^{2}/2\sigma^{2}(t',z)}.\label{eq:Gaussian bunch}
\end{equation}
The total current carried by the bunch of the profile \eqref{eq:Gaussian bunch}
is constant in any cross-section: $J=2\pi\sigma_{0}^{2}en_{0}v_{0}$. 

Let us assume, the bunch radius is modulated so that
\begin{eqnarray*}
\sigma(t',z) & = & \sigma_{0}\left[1+\epsilon f\left(k(z-v_{0}t')\right)\right]
\end{eqnarray*}
\begin{eqnarray}
 & = & \sigma_{0}\left[1+\epsilon f\left(kz-kv_{0}t+kR_{0}\frac{v_{0}}{c}-kr\frac{v_{0}}{c}\cos\varphi\right)\right]\label{eq:modulation}
\end{eqnarray}
and the modulation depth is small, $\epsilon\ll1$. Further, we assume
that the particle bunch is narrow, so that $k\sigma_{0}\ll1$. We
Taylor-expand \eqref{eq:Gaussian bunch}, keep terms linear in $\epsilon$
and quadratic in $kr$ and set the metal plate position at $z=-R_{0}v_{0}/c$:

\begin{equation}
n(t,\mathbf{r},z)\approx n_{0}\left[1+\epsilon A\left(\frac{r^{2}}{\sigma_{0}^{2}}-2\right)\right]e^{-r^{2}/2\sigma_{0}^{2}},\label{eq:Gaussian bunch-1}
\end{equation}
where
\begin{eqnarray*}
A & = & f\left(-kv_{0}t\right)-kr\frac{v_{0}}{c}f'\left(-kv_{0}t\right)\cos\varphi
\end{eqnarray*}
\begin{eqnarray}
 &  & +\frac{1}{2}\left(\frac{krv_{0}}{c}\right)^{2}f''\left(-kv_{0}t\right)\cos^{2}\varphi\label{eq:Gaussian taylor}
\end{eqnarray}
 The second and the third terms in \eqref{eq:Gaussian taylor} originate
from the time retardation \eqref{eq:retarded} and are essential for
the radiation emission.

The field excited by the axisymmetric particle bunch is then
\begin{eqnarray}
E(t,R_{0}) & = & -n_{0}\frac{2ev_{0}^{2}}{c^{2}R_{0}}\int_{0}^{2\pi}d\varphi\int_{0}^{\infty}r\times\nonumber \\
 &  & \left[1+\epsilon A\left(\frac{r^{2}}{\sigma_{0}^{2}}-2\right)\right]e^{-r^{2}/2\sigma_{0}^{2}}dr\label{eq:transverse-axisymmetric}
\end{eqnarray}

The integrand in \eqref{eq:transverse-axisymmetric} contains terms
with different powers of $kr$. The terms independent on $kr$ are
not retarded. The oscillatory part of the integral of these terms
vanishes because we assumed the total current of the particle bunch
being constant. These terms describe radiation at the wavelength defined
by the bunch current envelope. The terms linear in $kr$ contain the
factor $\cos\varphi$ and disappear when we integrate in the angle
$\varphi$ over the bunch cross-section. This is due to the axial
symmetry of current distribution in the bunch. The first oscillatory
terms that do not vanish after the integration are quadratic in $kr$.
Collecting the terms proportional to $\left(kr\right)^{2}$ at the
first time harmonic, we get
\begin{eqnarray*}
E(t,R_{0}) & = & \epsilon n_{0}f''\left(kv_{0}t\right)\frac{2ev_{0}^{4}}{R_{0}c^{4}}\int_{0}^{2\pi}\cos^{2}\varphi d\varphi\times
\end{eqnarray*}
\begin{eqnarray}
 & \times & \int_{0}^{\infty}r\left(kr\right)^{2}\left(1-\frac{r^{2}}{2\sigma_{0}^{2}}\right)e^{-\frac{r^{2}}{2\sigma_{0}^{2}}}dr.\label{eq:transverse-axisymmetric-1}
\end{eqnarray}
Integrating \eqref{eq:transverse-axisymmetric-1} we find the expression
for the radiation field at the distance $R_{0}$ from the beam axis:

\begin{equation}
E(t,R_{0})=-2\epsilon\frac{J}{cR_{0}}\left(k\sigma_{0}\right)^{2}\beta_{0}^{3}f''\left(-kv_{0}t\right).\label{eq:transverse-axisymmetric-general}
\end{equation}
The field is proportional to the radius modulation amplitude $\epsilon$,
square of the bunch radius $\left(k\sigma_{0}\right)^{2}$ and the
second derivative of the modulation function $f''\left(\omega_{rad}t\right)$.

In the case of a periodic harmonic modulation of the particle bunch
radius, $f(kv_{0}t)=\cos(kv_{0}t),$ the expression \eqref{eq:transverse-axisymmetric-general}
becomes 

\begin{equation}
E(t,R_{0})=2\epsilon\frac{J}{cR_{0}}\left(k\sigma_{0}\right)^{2}\beta_{0}^{3}\cos\omega_{rad}t,\label{eq:transverse-axisymmetric-2-1}
\end{equation}
where $J=2\pi\sigma_{0}^{2}en_{0}v_{0}$ is the total bunch current,
$\beta_{0}=v_{0}/c$ and the emitted radiation frequency is

\begin{equation}
\omega_{rad}=kv_{0}.\label{eq:frequency}
\end{equation}

We remember that the expression \eqref{eq:transverse-axisymmetric-general}
is valid only for narrow bunches, $k\sigma_{0}\ll1$. Wide particle
bunches, $k\sigma_{0}\gg1$, do not radiate coherently in the transverse
direction. The coherent radiated field becomes exponentially small
for large bunch radii. Thus, the optimal condition for TCTR is $k\sigma_{0}\sim1$.
This case is hard to treat analytically and we have to rely on numerical
simulations then. Still, the expression \eqref{eq:transverse-axisymmetric-general}
gives us a good guide to estimate the TCTR of an axially symmetric
proton bunch whose radius is modulated, but the total current in each
cross-section remains constant.

\begin{figure}
\includegraphics[width=0.9\columnwidth]{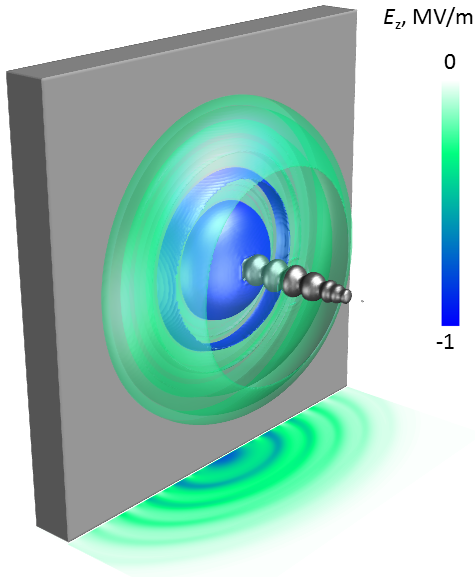}

\caption{\label{fig:modulated} TCTR emitted by a proton bunch exiting from
a metallic foil. Only the radius of the proton bunch is modulated,
not the current. The colored surfaces show the electric field component
normal to the metal foil.}
\end{figure}

As the proton beam in our case is modulated in plasma with some particular
density $n_{e}$, the modulation wave number $k$ is given by the
plasma wave number $k_{p}=\omega_{p}/c$, where $\omega_{p}=\sqrt{4\pi n_{e}e^{2}/m_{e}}$
is the plasma frequency. One can normalize the field \eqref{eq:transverse-axisymmetric-2-1}
on the so called plasma wave breaking field $E_{WB}=m_{e}c\omega_{p}/e$:

\begin{equation}
\frac{E(t,R_{0})}{E_{WB}}=\frac{2\epsilon}{kR_{0}}\frac{J}{J_{A}}\left(k\sigma_{0}\right)^{2}\beta_{0}^{3}\cos\omega_{rad}t,\label{eq:transverse-axisymmetric-normalized}
\end{equation}
where $J_{A}=mc^{3}/e=17$kA is the natural current value. The plasma
wave breaking field is given by the engineering formula $E_{WB}\approx\sqrt{n_{e}/10^{14}\mathrm{cm^{-3}}}$
GV/m. Although the formula \eqref{eq:transverse-axisymmetric-normalized}
has been derived for narrow weakly modulated bunches, it shows that
the radiated field strength may reach huge values for the available
proton bunches when these are deeply modulated, $\epsilon\sim1$,
and their radius is comparable with the modulation wavelength, $k\sigma_{0}\sim1$.

\section{TCTR emitted by hosed bunches with constant current}

Now we consider TCTR of a particle bunch whose radius is constant,
while the centroid position oscillates periodically in a plane. We
assume the Gaussian transverse bunch density distribution
\begin{equation}
n(t',\mathbf{r},z)=n_{0}e^{-\left(\mathbf{r}-\mathbf{r}_{c}(t',z)\right)^{2}/2\sigma^{2}}.\label{eq:hosing}
\end{equation}
Again, the bunch is assumed to be thin, $k\sigma\ll1$. The total
current carried by the bunch of the profile \eqref{eq:hosing} is
constant in any cross-section: $J=2\pi\sigma^{2}en_{0}v_{0}$. 

For the bunch centroid displacement we write
\begin{eqnarray*}
\mathbf{r}_{c}(t',z) & = & \mathbf{r}_{c0}h\left(k(z-v_{0}t')\right)
\end{eqnarray*}
\begin{eqnarray}
 & = & \mathbf{r}_{c0}h\left(kz-kv_{0}t+kR_{0}v_{0}/c-kr(v_{0}/c)\cos\varphi\right)\label{eq:hosing radius}
\end{eqnarray}
where the bunch centroid displacement is small, $|\mathbf{r}_{c0}|/\sigma\ll1$.
We Taylor-expand \eqref{eq:hosing}, keep terms linear in $\epsilon$
and $kr$ and set the metal plate position at $z=-R_{0}v_{0}/c$:

\begin{equation}
n(t,\mathbf{r},z)\approx n_{0}\left(1+\frac{\mathbf{r}_{c0}\mathbf{r}}{\sigma^{2}}A_{h}\right)e^{-r^{2}/2\sigma_{0}^{2}},\label{eq:hosed bunch}
\end{equation}
where 
\begin{equation}
A_{h}=h\left(-kv_{0}t\right)-kr\frac{v_{0}}{c}h'\left(-kv_{0}\right)t\cos\varphi\label{eq:hosed taylor}
\end{equation}
and 

\begin{equation}
\mathbf{r}_{c0}\mathbf{r}=r_{c0}r\cos(\varphi-\varphi_{0})
\end{equation}
with $\varphi_{0}$ being the angle between the plane of hosing and
the direction to the observer. 

Because the hosed bunch is not axisymmetric, it is sufficient to keep
the linear terms in $kr$ only. The second term in \eqref{eq:hosed taylor}
originates again from the time retardation \eqref{eq:retarded} and
it is this term that leads to the radiation emission.

The field excited by the axisymmetric particle bunch is then
\begin{eqnarray*}
E(t,R_{0}) & = & -n_{0}\frac{2ev_{0}^{2}}{c^{2}R_{0}}\int_{0}^{2\pi}d\varphi\int_{0}^{\infty}r\times
\end{eqnarray*}
\begin{eqnarray}
 & \times & \left(1+\frac{r_{c0}r\cos(\varphi-\varphi_{0})}{\sigma^{2}}A_{h}\right)e^{-\frac{r^{2}}{2\sigma_{0}^{2}}}dr.\label{eq:transverse-hosed}
\end{eqnarray}
In the integrand in \eqref{eq:transverse-hosed}, we collect terms
linear in $kr$. The terms independent on $kr$ are not retarded.
The oscillatory part of the integral of these terms vanishes because
we assumed the total current of the particle bunch being constant.
The terms, linear in $kr$ contain the factor $\cos\varphi$ and do
not disappear in general. Collecting the terms proportional to $kr$
and integrating we get
\begin{eqnarray*}
E(t,R_{0}) & = & 2\pi\cos\varphi_{0}h'\left(-kv_{0}t\right)\frac{en_{0}v_{0}^{3}}{R_{0}c^{3}}\sigma^{2}kr_{0}
\end{eqnarray*}
\begin{eqnarray}
 & = & \frac{J}{cR_{0}}\beta_{0}^{2}kr_{0}\cos\varphi_{0}h'\left(-kv_{0}t\right)\label{eq:transverse-hosed-field}
\end{eqnarray}
 The radiation amplitude is proportional to the hosing amplitude $kr_{0}$
and the first derivative of the hosing function $h'\left(\omega_{rad}t\right)$.
In the case of a periodic harmonic hosing of the particle bunch radius,
$h(kz)=\cos(kz),$ the expression \eqref{eq:transverse-hosed-field}
becomes 
\begin{equation}
E(t,R_{0})=\frac{J}{cR_{0}}\beta_{0}^{2}kr_{0}\cos\varphi_{0}\sin\left(kv_{0}t\right)\label{eq:field hosed harmonic}
\end{equation}
From \eqref{eq:transverse-hosed-field}-\eqref{eq:field hosed harmonic}
we see that the hosed bunch radiation is not axially symmetric. The
main lobes are located in the hosing plane as described by the factor
$\cos\varphi_{0}$ and are antisymmetric to each other.

\begin{figure}
\includegraphics[width=0.9\columnwidth]{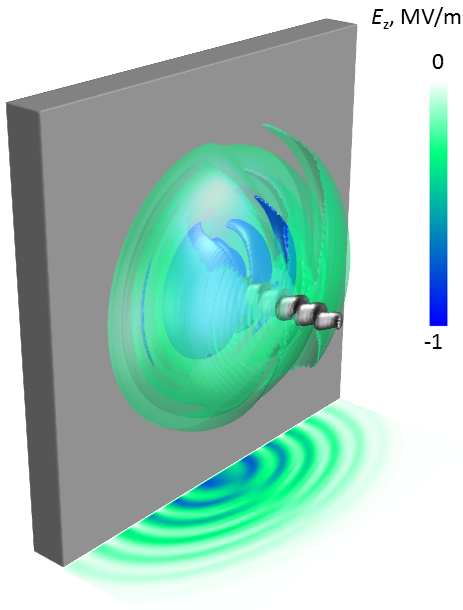}

\caption{\label{fig:hosed} TCTR emitted by a hosed proton bunch exiting from
a metallic foil. The radiated field is shaken off anti-symmetrically
in the plane of hosing.}
\end{figure}

Again, one can normalize the field \eqref{eq:field hosed harmonic}
on the plasma wave breaking field $E_{WB}=m_{e}c\omega_{p}/e$ if
the proton bunch has been hosed during its propagation in plasma:

\begin{equation}
\frac{E(t,R_{0})}{E_{WB}}=\frac{1}{kR_{0}}\frac{J}{J_{A}}\beta_{0}^{2}kr_{0}\cos\varphi_{0}\sin\left(kv_{0}t\right)\label{eq:transverse-hosed-normalized}
\end{equation}
The radiated field \eqref{eq:transverse-hosed-normalized} scales
linearly with the hosing amplitude $kr_{0}$.

\section{Numerical examples}

We performed two test numerical simulations to illustrate the effect
of TCTR. The simulations have been done with the hybrid code H-VLPL3D
\cite{h-vlpl3d,vlpl}. The code can handle plasmas of arbitrary high
densities as fluids. In these test simulations, we take a model proton
bunch that carries the current of an optimized Super-Proton-Synchrotron
(SPS) bunch, $I_{0}=50$A and the protons have the longitudinal momentum
$p=450$GeV/c. The length of the model bunch was chosen short, $\sigma_{z}=2.4$mm,
for illustration purposes. The final result does not depend on the
bunch length.

In the first simulation of a modulated proton bunch, we set up the
model density profile

\begin{equation}
n(t,\mathbf{r},z)=n_{0}\frac{\sigma_{0}^{2}}{\sigma^{2}(t',z)}e^{-r^{2}/2\sigma^{2}(t,z)}e^{-z^{2}/2\sigma_{z}^{2}}.\label{eq:test modulated}
\end{equation}
with $n_{0}=1.4\cdot10^{12}\mathrm{cm^{-3}}$ and the unperturbed
transverse radius $\sigma_{0}=0.2$mm. The actual bunch radius is
modulated as $\sigma=\sigma_{0}\left(1+0.5\cos k_{p}z\right)$, where
$k_{p}=2\pi/\lambda_{p}$ is the plasma wave number. For the plasma
density of $n_{e}=7\cdot10^{14}\mathrm{cm^{-3}}$ we have the plasma
wavelength $\lambda_{p}=1.2$ mm. 

In the simulation, we send the proton bunch through a high density
plasma slab that would represent a conducting metal foil. The foil,
the proton bunch and the radiated field are shown in Fig. \ref{fig:modulated}.
The radiation is emitted axisymmetrically in the transverse direction.
The radiation field magnitude reaches several 100kV/m values at 1mm
distance from the bunch axis.

In the second simulation we used a hosed proton bunch with the model
density profile

\begin{equation}
n(t,\mathbf{r},z)=n_{0}e^{-(x-x_{0}(t,z))^{2}/2\sigma_{0}^{2}}e^{-z^{2}/2\sigma_{z}^{2}}.\label{eq:test modulated-1}
\end{equation}
with $n_{0}=1.4\cdot10^{12}\mathrm{cm^{-3}}$,$\sigma_{0}=0.2$mm
and the time dependent centroid $x_{0}=r_{0}\cos k_{p}z$, where the
hosing amplitude was $r_{0}=0.2$mm. The radiation produced by this
proton bunch as it traverses a metal foil is shown in Fig. \ref{fig:modulated}.
The radiation is no more axisymmetric. Rather, the hosed bunch {}``shakes
off'' its field in the direction of hosing. The radiation field again
reaches 100 kV/m values at 1mm distance from the bunch axis.

\section{Conclusions}

It has been shown that a particle bunch that is modulated in radius
or hosed does radiate TCTR even when the total current of the bunch
remains unmodulated. The radiation is emitted due to the retardation
effect: the observer located in the transverse plane sees signals
of particles traversing the conductor plate at retarded times. The
integral of this retarded current does show a time dependance at the
bunch modulation period. An axisymmetric particle bunch with a periodically
modulated radius emits axisymmetric waves. On the contrary, a hosed
bunch whose centroid oscillates in a transverse plane emits in a direction
perpendicular to this plane and does not emit in the plane of hosing.
In both cases, the field of the coherent radiation scales linearly
with the bunch current and is proportional to the modulation amplitude.
The TCTR is emitted only when the bunch radius is smaller than or
comparable with the radiation wavelength. The emission is maximized
when the bunch radius is $k\sigma\sim1$, because in this case the
retardation effects maximize while the coherence is not yet affected
by the finite bunch radius. The numerical simulations suggest that
plasma-modulated SPS proton bunches may emit TCTR fields at the characteristic
amplitudes of 100 kV/m.

This work was supported by DFG.

\end{document}